\documentclass[prl,aps,amsmath,twocolumn,floatfix,showpacs]{revtex4}
\usepackage{epsfig}

\begin{document}

\title{Precision measurement and compensation of
optical Stark shifts \newline for an ion-trap quantum processor}

\author{H.~H\"affner, S.~Gulde, M.~Riebe, G.~Lancaster, C.~Becher, J.~Eschner,
F.~Schmidt-Kaler, and R.~Blatt}

\affiliation{Institut f{\"u}r Experimentalphysik, 6020 Innsbruck, Austria}

\date{\today}

\begin{abstract}
Using optical Ramsey interferometry, we precisely measure the laser-induced
AC-stark shift on the $\textrm{S}_{1/2}$--$\textrm{D}_{5/2}$ ``quantum bit''
transition near 729~nm in a single trapped $^{40}$Ca$^+$ ion. We cancel this
shift using an additional laser field. This technique is of particular
importance for the implementation of quantum information processing with cold
trapped ions. As a simple application we measure the atomic phase evolution
during a $n \times 2\pi$ rotation of the quantum bit.

\end{abstract}

\pacs{42.50.Hz 39.20.+q 32.80.Qk}


\maketitle







Atomic two-level systems are currently discussed extensively for quantum
information processing \cite{CHUANG00,PhysQI00,SASURA02}. Typically, a qubit
is encoded in an atomic transition between levels with extremely slow
radiative decay: Hyperfine ground states are used as qubits which are
connected via a far-off-resonant Raman transition \cite{WINE98}.
Alternatively, the qubit is encoded in a superposition of a ground state and
a long-lived metastable state \cite{ROOS99}, manipulated on a direct optical
quadrupole transition. In both cases the transition matrix element between
the qubit levels is small, such that relatively high laser intensities are
required to drive the transition. These strong electromagnetic fields will
unavoidably lead to level shifts known as the dynamical or AC-Stark shift
\cite{DRAKEAMO}.

Even though a pure two-level system does not show AC-Stark shift if driven by
resonant laser radiation, in real physical situations additional energy
levels exist which lead to significant AC-Stark shifts of the qubit energy
levels. Since quantum algorithms manifest themselves in many-particle quantum
interference, any uncontrolled phase shift induced by manipulation of the
qubits complicates the implementation of these algorithms considerably and
must be avoided. The interest in fast one- and two-qubit manipulations
aggravates the problem \cite{STEANE00}, since strong laser fields are
required. Therefore complications due to AC-Stark shifts arise in most
realizations of quantum information processing with trapped ions
\cite{CIRAC95,SOERENSEN99}. Similar problems exist in precision spectroscopy
\cite{WEISS94} and with optical clocks \cite{SENGSTOCK94,NIERING00}.

In this paper we present a measurement which identifies different
contributions to phase shifts of a qubit caused by AC-Stark level shifts. We
determine the relative oscillator strengths of various contributing
transitions. Finally, we show how to compensate for the AC-Stark shift
experimentally. This compensation greatly simplifies the implementation of
quantum algorithms. As an application of the compensation method we
demonstrate the controlled sign change of a qubit wavefunction under rotation
by 2$\pi$.

\begin{figure}[htb]
\begin{center}
\epsfig{file=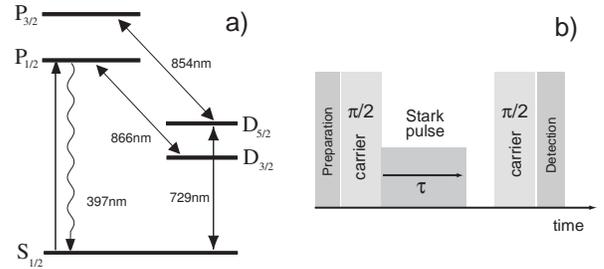,width=.9\linewidth} \caption{\label{prinzip}
a) Level scheme of Ca$^+$ ion. b) Ramsey method to detect the
AC-Stark effect. The $\pi/2$ pulses are resonant, the Stark pulse
is detuned. See text for more details.}
\end{center}
\end{figure}

In the following we focus on the qubit transition between the S$_{1/2}$
ground state and the metastable D$_{5/2}$ state (lifetime $\simeq$~1~s) in
$^{40}$Ca$^+$, see Fig.~\ref{prinzip}a. We drive coherently the transition
between the Zeeman components $|\textrm{S}_{1/2}, m=-\frac{1}{2} \rangle
\equiv |\textrm{S}\rangle$ and $|\textrm{D}_{5/2}, m'=-\frac{1}{2} \rangle
\equiv |\textrm{D}\rangle$. AC-Stark shifts of these levels are induced by
off-resonant quadrupole coupling to other transitions within the
$\textrm{S}_{1/2}-\textrm{D}_{5/2}$ Zeeman manifold, as well as by
off-resonant dipole coupling to other levels
($\textrm{S}_{1/2}-\textrm{P}_{1/2},~ \textrm{S}_{1/2}-\textrm{P}_{3/2},~
\textrm{D}_{5/2}-\textrm{P}_{3/2}$).

We measure the AC-Stark shift induced by a laser pulse detuned from the
$|\textrm{S}\rangle - |\textrm{D}\rangle$ transition (``Stark pulse'') with a
Ramsey interference experiment \cite{RAMSEY85}. For this, we enclose the
Stark pulse within a pair of Ramsey $\pi/2$ pulses resonant with the
$|\textrm{S}\rangle - |\textrm{D}\rangle$ transition (see
Fig.~\ref{prinzip}b). Starting from the ground state $|\textrm{S}\rangle$,
the first resonant Ramsey pulse transfers the ion to the superposition
$\Psi_{1} = \{|\textrm{S}\rangle + i |\textrm{D}\rangle\} /\sqrt{2}$
(rotating frame). The Stark pulse shifts the levels $|\textrm{S}\rangle$ and
$|\textrm{D}\rangle$ for the time $\tau$ it is applied. Since the coupling of
this pulse to both levels is different, we denote the phases which are
acquired as $\phi_S=\tau \delta_S$ and $\phi_D=\tau \delta_D$, respectively,
and define $\Delta\phi=\phi_D-\phi_S$. Immediately before the second Ramsey
pulse, the wave function is $\Psi_{2} = \{e^{-i\tau\delta_{S}}
|\textrm{S}\rangle + i e^{-i\tau\delta_{D}} |\textrm{D}\rangle\} /\sqrt{2}$.
Now the second Ramsey pulse is applied, transferring phase information into
population and yielding the state $\Psi_{3} = \frac{1}{2} e^{-i\phi_S} \{
(1-e^{-i\Delta\phi}) |\textrm{S}\rangle + i(1+e^{-i\Delta\phi})
|\textrm{D}\rangle \}$. Finally, $\Psi_{3}$ is projected by a measurement of
the probability to find the ion in the state $|\textrm{D}\rangle$, $P_D =
\frac{1}{2}(1+\cos(\Delta\phi))$. With this method we obtain the phase shift
$\Delta\phi$ induced by a Stark pulse applied for duration $\tau$. For a
systematic measurement we vary the Stark pulse duration $\tau$ while keeping
the separation of the Ramsey pulses constant. The frequency of the population
variation $P_D(\tau)$ directly yields the Stark shift $\delta_D-\delta_S
\equiv \delta_{{\rm AC}}$.

For the experiments, a single $^{40}$Ca$^+$ ion is stored in a linear Paul
trap made of four blades separated by 2~mm for radial confinement and two
tips separated by 5~mm for axial confinement.  Under typical operating
conditions we observe axial and radial motional frequencies $(\omega_{ax},
\omega_r)/2\pi =$~(1.7, 5.0)~MHz. The trapped $^{40}$Ca$^+$ ion has a single
valence electron and no hyperfine structure.

We perform Doppler cooling of the ion on the $\textrm{S}_{1/2} -
\textrm{P}_{1/2}$ transition at 397~nm. Diode lasers at 866~nm and 854~nm
prevent pumping into the D states. For sideband cooling and for quantum
processing \cite{ROOS99}, we excite the S$_{1/2}$ to D$_{5/2}$ transition
with a Ti:Sapphire laser near 729~nm ($\leq 100$~Hz linewidth). With
$\simeq$~30~$\mu$m beam waist diameter and $\simeq$~50~mW laser power we
achieve Rabi frequencies around 1~MHz, measured by driving
$|\textrm{S}\rangle$ to $|\textrm{D}\rangle$ Rabi oscillations  resonantly
\cite{WINE98,ROOS99}. A constant magnetic field of 2.4~G splits the 10 Zeeman
components of the S$_{1/2}$ to D$_{5/2}$ multiplet. The chosen geometry and
polarization allow excitation of $\Delta m = 0$ and $\pm~2$ transitions only.
We detect whether a transition to D$_{5/2}$ has occurred by applying the
laser beams at 397~nm and 866~nm and monitoring the fluorescence of the ion
on a photomultiplier (electron shelving technique). The internal state of the
ion is discriminated with an efficiency close to 100$\%$ within 3~ms
\cite{ROOS99}.

The measurement cycle (total duration 20~ms) consists of four consecutive
steps: (i) Doppler cooling leads to low thermal vibrational states of axial
and radial modes, $\langle n_{r} \rangle \approx 3$. (ii) Sideband cooling of
the axial motion is performed on the $|\textrm{S}_{1/2}, m=-\frac{1}{2}
\rangle \leftrightarrow |\textrm{D}_{5/2}, m'=-\frac{5}{2} \rangle$
transition, leading to more than 99~$\%$ ground state population. Pumping
into $|\textrm{S}_{1/2}, m=+\frac{1}{2} \rangle$ is counteracted by several
short pulses of $\sigma^-$ radiation at 397~nm. (iii) Ramsey spectroscopy on
the $|\textrm{S}\rangle$ to $|\textrm{D}\rangle$ transition, see
Fig.~\ref{prinzip}b. The $\pi/2$ pulses have 8~$\mu$s duration and 200~$\mu$s
separation. An intermediate off-resonant laser pulse (Stark pulse) with
duration $\tau$, Rabi frequency $\Omega_{{\rm AC}}$ and detuning
$\Delta_{{\rm AC}}$ induces a phase shift $\Delta\phi = \delta_{\rm AC}\tau$.
(iv) The final state after the second Ramsey pulse is determined by electron
shelving. After this, laser light at 854~nm pumps the ion back to the
S$_{1/2}$ ground state.
\begin{figure}[ht]
\begin{center}
\epsfig{file=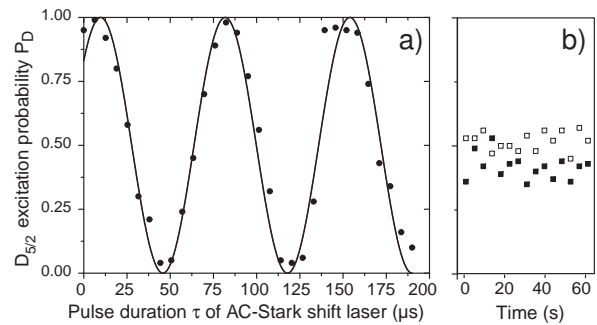,width=.9\linewidth} \caption{\label{sinus} (a) Ramsey
pattern: The AC-Stark shift is determined as the oscillation frequency of
$P_D(\tau)$, here $\delta_{{\rm AC}}/2\pi$ = 13.9(0.2)~kHz. The phase offset
is due to a small detuning of the Ramsey pulses. (b) Compensation of the
AC-Stark effect: The ion is illuminated by the Stark pulse and an additional
off-resonant compensation laser field which causes an equal AC-Stark shift,
but with opposite sign. Data are taken alternating, with $\tau=0~\mu$s
(black) and $200~\mu$s (white), see text for details. We estimate a residual
$\delta_{{\rm AC}}/2\pi =$~0.25(3)~kHz.}
\end{center}
\end{figure}

\begin{figure}[b]
\begin{center}
\epsfig{file=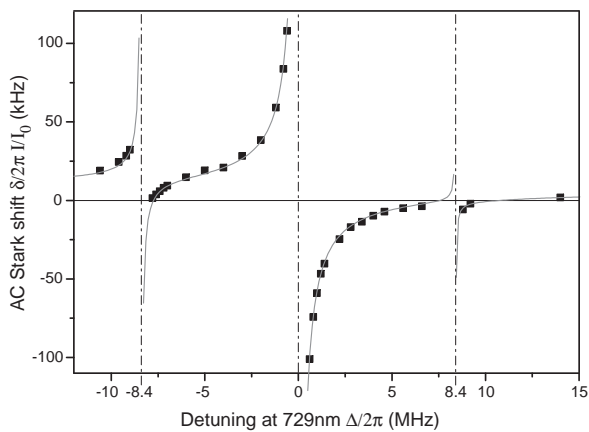,width=.9\linewidth} \caption{\label{StarkSpekSchmal}
The measured ac-Stark shift data (see Fig.~\ref{sinus}a) are normalized
according the measured laser power $I(\Delta)/I(0)$ which varies by about
50~\% over the whole tuning range of $\Delta$. This normalized data (squares)
and calculated (line) Stark shift $\delta_{{\rm AC}}$ (Eq.~ref) are plotted
versus the detuning $\Delta_{{\rm AC}}$ of the Stark pulse from the
$|\textrm{S}\rangle - |\textrm{D}\rangle$ resonance. The divergences are due
to the $(m=-\frac{1}{2}) \leftrightarrow (m^\prime = -\frac{5}{2},
-\frac{1}{2}, +\frac{3}{2})$ resonances (from left to right). Two data points
at large detunings are not shown. They read $\delta_{{\rm AC}}/2\pi =
5.88$~kHz and 8.49~kHz for detunings $\Delta_{{\rm AC}}/2\pi = 40$~MHz and
60~MHz, respectively, and are equally well described by the theoretical
curve.}
\end{center}
\end{figure}

The sequence (i)-(v) is repeated 100~times to measure $P_D$ for
each value of $\tau$ and $\Delta_{\rm AC}$. Varying $\tau$ yields
data as shown in Fig.~\ref{sinus}a. From the fit of $P_D$ to the
data we find the AC-Stark shift $\delta_{{\rm AC}}$. For a given
$\Delta_{{\rm AC}}$ each experiment was conducted twice to cancel
a slow phase drift of the Ramsey pulses due to a laser frequency
drift on the order of 1~Hz/s: First the Stark pulse duration was
increased with each step from $\tau = 0$ to $\tau = 200~\mu$s,
then it was decreased again. The frequency detuning $\Delta_{{\rm
AC}}$ of the Stark pulse is varied over several Zeeman resonances
of the $\textrm{S}_{1/2} - \textrm{D}_{5/2}$ manifold. Note that
we extract only the modulus but not the sign of $\delta_{{\rm
AC}}$ from the fit of $P_D$. The signs are attributed to the
measured results according to the theoretical expectation
discussed below. We determine the variation of the light intensity
$I(\Delta)$ with the laser frequency detuning using a calibrated
powermeter. From this we normalize the ac-Stark shift as
$\delta_{{\rm AC}} I(\Delta)/I(0)$ to obtain the data plotted in
Fig.~\ref{StarkSpekSchmal}.

There are three contributions to the AC-Stark shift: In our parameter regime
the largest contribution is due to the different Zeeman transitions permitted
by our particular geometry ($m=-\frac{1}{2}$ to $m'=\frac{3}{2},
-\frac{1}{2},$ and $-\frac{5}{2}$). The second largest contribution arises
from off-resonant dipole coupling of S$_{1/2}$ to P$_{1/2}$, P$_{3/2}$, and
from D$_{5/2}$ to P$_{3/2}$. A third contribution is caused by motional
sidebands: For a trapped ion, the Rabi frequencies on the red and blue
motional sideband are given by $\Omega_{\rm SB}=\Omega \eta \sqrt{n}$ and
$\Omega \eta \sqrt{n+1}$, respectively, where $\eta$ denotes the Lamb-Dicke
factor and $n$ the phonon quantum number in the vibrational mode. With an ion
cooled to the ground state of motion ($\langle n_{ax} \rangle \approx 0$) and
with $\eta_{ax} = 0.07$, we expect $\leq 0.5\%$ relative contribution of the
blue axial sideband to the overall AC-Stark shift. The contribution of the
red axial sideband vanishes. Similar reasoning holds for the radial motion
with $\langle n_{rad} \rangle \approx 3$ and $\eta_{rad}=0.016$. We therefore
neglect these contributions in the following.

The other relevant contributions to the AC-Stark shift can be summarized as:
\begin{equation}\label{TotShift}
\delta_{{\rm AC}} = \frac{\Omega^2_{\rm AC}}{4} \left( 2b-
\frac{a_{-5/2}}{\Delta_{\rm AC}-\Delta_{-5/2}} -2\frac{a_{-1/2}}{\Delta_{\rm
AC}}- \frac{a_{3/2}}{\Delta_{\rm AC}-\Delta_{3/2}} \right)\;.
\end{equation}
Here $\Omega_{\rm AC}$ denotes the Rabi frequency of the Stark laser field.
The transitions $m=-\frac{1}{2} \leftrightarrow m'=-\frac{5}{2},
+\frac{3}{2}$ are at $\Delta_{-5/2}$ and $\Delta_{+3/2}$, which are $\pm
(2\pi)8.4$~MHz in the experiment. An explicit calculation of the matrix
elements of quadrupole transitions is given in Refs.~\cite{JAMES98,ROOS_PHD}.
The coefficients $a_{-1/2}$, $a_{-5/2}$ and $a_{3/2}$ are the squares of the
relative coupling strengths. We define $a_{-1/2}$ to be one, since the Ramsey
spectroscopy is carried out on this transition. From the laser polarization
and laser axis with respect to the magnetic field axis we calculate
\cite{ROOS_PHD} $a_{-5/2}=0.278$ and $a_{3/2}=0.0556$. The factor of 2 in the
contribution of the $m=-\frac{1}{2}$ to $m'=-\frac{1}{2}$ transition appears
because the Ramsey method is applied on this transition such that the shift
of both the upper and lower state is detected. From the other Zeeman
components, however, only the shift of the lower state $|\textrm{S}\rangle$
becomes apparent. The constant $b$ in Eq.~(\ref{TotShift}) contains the
squared relative coupling strengths to all other dipole transitions. No
dependence on the laser detuning appears since the transitions are far
off-resonant.

The optimum fit of Eq.~(\ref{TotShift}) to the data in
Fig.~\ref{StarkSpekSchmal} is obtained with $a_{-5/2}=0.32\,(2)$,
$a_{+3/2}=0.05\,(2)$, $b=0.112\,(5)/2\pi~($MHz$)^{-1}$ and $\Omega_{{\rm
AC}}/2\pi= 357\,(3)$~kHz. We independently measured $a_{-5/2}= 0.36\,(2)$ and
$a_{+3/2}= 0.05\,(1)$  with resonant Rabi oscillations. These values agree
within their error margins with those obtained from the fit to the Stark
shift data.

Most of the current proposals for quantum computation require that the ion is
driven on the motinal sidebands. Applying a laser on the blue axial sideband
of the $|\textrm{S}\rangle \leftrightarrow |\textrm{D}\rangle$ transition
(``gate laser'') at a detuning of $\Delta/2\pi$= +1.7~MHz results in a
negative AC-Stark shift. However, shining in a second light field at a
frequency whose AC-Stark shift is positive can compensate for this unwanted
phase shift.
As discussed in the introduction, such an AC-Stark shift cancellation is a
prerequisite for any quantum algorithm. Our method to determine the optimum
setting of the compensation laser consists of the following steps: First we
detune the gate laser by $\simeq$~80~kHz from the sideband resonance to avoid
excitation into the D state (its AC-Stark effect however is still practically
identical to that of a laser field resonant with the sideband). Then we
minimize the total AC-Stark effect by adjusting the intensity and detuning of
the compensation laser field such that the oscillations in $P_D$ disappear.
Both light fields are generated from the output of a single Ti:Sapphire laser
by driving a broad-band AOM (in double pass configuration) with the two
rf-frequencies simultaneously. Since both light fields are derived from the
same laser, intensity fluctuations do not affect the compensation.

The accuracy to which the AC-Stark effect can be nulled is proportional to
$(2~T_R)^{-1}~S/N $, where $T_R$ denotes the Ramsey interrogation time (here
$200~\mu$s) and $S/N$ the signal to noise ratio of the state measurement.
Integrating this measurement for long times, to improve $S/N$, is limited by
the frequency drift of the laser source near 729~nm (typically $\leq$
1~Hz/s), since a drift of the relative phase of the Ramsey pulses mimics a
residual AC-Stark effect. To overcome this problem, we optimize the Rabi
frequency and detuning of the compensation laser by alternating Ramsey
experimental sequences with $\tau=0$ and $\tau \simeq 200 \mu$s. Thus, a slow
drift is discriminated against a residual phase shift due to imperfect
compensation. Limited by the shot noise of $P_D(\tau)$, any AC-Stark effect
can be cancelled to within $\simeq ~2\pi$30~Hz in 60~s. See Fig.~\ref{sinus}b
for the compensation data over the course of time, each data point
corresponding to 100 repetitions of the experimental sequence and a time
duration of 2~s.

As an application of the compensation method we demonstrate the sign change
of a wavefunction, a simple building block frequently required in quantum
algorithms. A driven spin-1/2-system transforms into its initial state only
by a $4\pi$ rotation, whereas a $2\pi$ rotation leads to a sign change of the
wavefunction. This phase shift is the central part of the Cirac--Zoller
proposal \cite{CIRAC95} for quantum gates with trapped ions \cite{MONROE95}.
Similarly, Ramsey experiments on Rydberg atoms have been performed in the
microwave regime, to investigate the AC-Stark shift of the electromagnetic
vacuum field \cite{BRUNE94} and to perform a tunable phase gate
\cite{RAUSCH99}.
\begin{figure}[htb]
\begin{center}
\epsfig{file=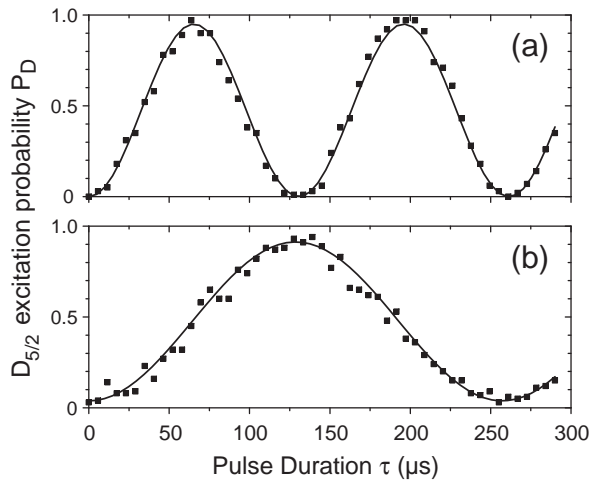,width=.9\linewidth} \caption{\label{2pi4pi} (a)
Resonant Rabi oscillations on the blue sideband of the $|\textrm{S}\rangle -
|\textrm{D}\rangle$ transition. The period of the population oscillation is
131(1)~$\mu$s, as found from the fit to the data. (b) Ramsey $\pi$/2 pulses
on the $|\textrm{S}\rangle - |\textrm{D}\rangle$ carrier transition enclose
the Rabi flopping on the sideband. The phase of the $|\textrm{S}\rangle$
state is revealed to oscillate with a period of 257(2)~$\mu$s. The ratio of
both periods is 1.96(3) and agrees well with the expected value of 2.}
\end{center}
\end{figure}

In our experiment, the ion is first prepared in the vibrational and
electronic ground state and then driven resonantly between the
$|\textrm{S},n=0\rangle$ and $|\textrm{D},n=1\rangle$ state (blue axial
sideband near $\Delta$=+1.7~MHz), with the AC-Stark compensation laser field
switched on. Fig.~\ref{2pi4pi}a shows the corresponding Rabi oscillations.

To measure the phase acquired during the sideband interaction we
enclosed the Rabi oscillations between two carrier
($\Delta=0$~MHz) Ramsey $\pi/2$-pulses with phases 0 and $\pi$.
Under the first Ramsey pulse the initial state
$|\textrm{S}\rangle$ transforms into $\left(|\textrm{S}\rangle +
i|\textrm{D}\rangle\right) / \sqrt{2}$, and is rotated back into
$|\textrm{S}\rangle$, in case of zero sideband interaction time.
If, however, the time for the sideband interaction corresponds to
a $2\pi$ rotation on the blue sideband, the acquired phase of $-1$
of the population in the $|\textrm{S}\rangle$-state results in the
state $\left(-|\textrm{S}\rangle\ + i|\textrm{D}\rangle\right) /
\sqrt{2}$, which is transformed to $-i|\textrm{D}\rangle$ by the
second Ramsey pulse. Only the 4$\pi$ rotation of the qubit leads
back to the initial state. The experimental finding is presented
in Fig.~\ref{2pi4pi}: The $2\pi$ rotation, near 131~$\mu$s
interaction time, shows up as a -1 phase shift, while after about
260~$\mu$s a full $4\pi$ rotation is completed. In this example,
the compensation laser corrected for an AC-Stark shift of
$\delta_{{\rm AC}}\approx~2\pi\cdot3.1$~kHz. Without compensation,
this shift alone would have resulted in an additional phase of
$\sim$0.82$\pi$ in 131~$\mu$s.

In conclusion we have precisely measured optical AC-Stark shifts
on a single ion using an optical Ramsey interferometer. The method
highlights the phase evolution an atom undergoes while it is
irradiated off-resonantly with light. Measuring on a single ion
removes the influence of the beam intensity profile on the
measurement result, typically occuring with atom samples. We have
also demonstrated how to compensate for the AC-Stark shift with a
simultaneous laser pulse at another frequency. The compensation
enables quantum computing on optical transitions and to detect
phases of multi-particle quantum interference in a straightforward
way. The quality to which the AC-Stark effect can be cancelled is
proportional to the chosen Ramsey interrogation time. Therefore,
it is likely that the quality of this novel phase compensation
method will improve when in future longer coherence times can be
realized, as necessary for a large number of quantum gate
operations.

\acknowledgments{This work is supported by the Austrian 'Fonds zur
F\"orderung der wissenschaftlichen Forschung' SFB15, by the
European Commission (QSTRUCT and QI networks, ERB-FMRX-CT96-0077
and -0087, QUEST network, HPRN-CT-2000-00121, QUBITS network,
IST-1999-13021), and by the "Institut f\"ur Quanteninformation
GmbH".}



\end{document}